\documentclass[prl,twocolumn,superscriptaddress]{revtex4}
\usepackage{epsfig}
\usepackage{times}
\usepackage{amsmath}
\usepackage{amsfonts}
\usepackage{amssymb}
\usepackage[usenames]{color}
\usepackage{graphicx}

\newcommand{\beq}{\begin{equation}}
\newcommand{\eeq}{\end{equation}}
\newcommand{\beqa}{\begin{eqnarray}}
\newcommand{\eeqa}{\end{eqnarray}}

\newcommand{\NTT}{NTT Basic Research Laboratories, NTT Corporation, 3-1
Morinosato-Wakamiya, Atsugi, Kanagawa, 243-0198, Japan.}
\newcommand{\china}{Institute of Physics, Chinese Academy of Sciences,
Beijing, 100190, China.}

\newcommand{\OSAKA}{Graduate School of Engineering Science, University of Osaka, 1-3 Machikane-yama, Toyonaka, Osaka 560-8531, Japan.}
\newcommand{\NII}{National Institute of Informatics, 2-1-2 Hitotsubashi,
Chiyoda-ku, Tokyo 101-8430, Japan.}
\newcommand{\NICT}{National Institute of Information and Communications Technology, 4-2-1, Nukuikitamachi, Koganei-city, Tokyo 184-8795 Japan}
\begin{document}
\title{Improving the lifetime of the NV center ensemble coupled with a
superconducting flux qubit by applying magnetic
fields}

\author{\ \ \ \ \ \ \ \ \ \ \ \ \ \ \ \ \ \ \ \ \ Yuichiro Matsuzaki}			\email{matsuzaki.yuichiro@lab.ntt.co.jp} \affiliation{\NTT}
\author{Xiaobo Zhu}		\affiliation{\china}
\author{Kosuke Kakuyanagi}			\affiliation{\NTT}
\author{Hiraku Toida}			\affiliation{\NTT}
\author{Takaaki Shimooka}			\affiliation{\OSAKA}
\author{\newline Norikazu Mizuochi}			\affiliation{\OSAKA}
\author{ Kae Nemoto}				\affiliation{\NII}
\author{Kouichi Semba}				\affiliation{\NICT}
\author{W. J. Munro}				\affiliation{\NTT}
\author{Hiroshi Yamaguchi} 	\affiliation{\NTT}
\author{Shiro Saito} 	\affiliation{\NTT}
\begin{abstract}
One of the promising systems to realize quantum computation
 is a hybrid system
 where a superconducting flux qubit plays a
 role of a quantum processor and the NV center ensemble is used as a
 quantum memory. 
 \textcolor{black}{
We have theoretically and experimentally studied the effect of
 magnetic fields
 on this hybrid system,
 and found that the
 lifetime of the vacuum Rabi oscillation
 is improved by
 applying a few mT magnetic field to the NV center ensemble.}
 Here, we construct a theoretical model to
 reproduce the vacuum Rabi \textcolor{black}{oscillations} with/without magnetic fields applied to the
 NV centers, and we determine the reason why magnetic fields can
 affect the coherent properties of the NV center ensemble. From our
 theoretical analysis, we quantitatively show that the magnetic fields
 actually suppress the inhomogeneous broadening \textcolor{black}{from} the strain in the
 NV$^-$ centers.
\end{abstract}

\maketitle

\begin{figure}[ht]
\includegraphics[scale=0.17]{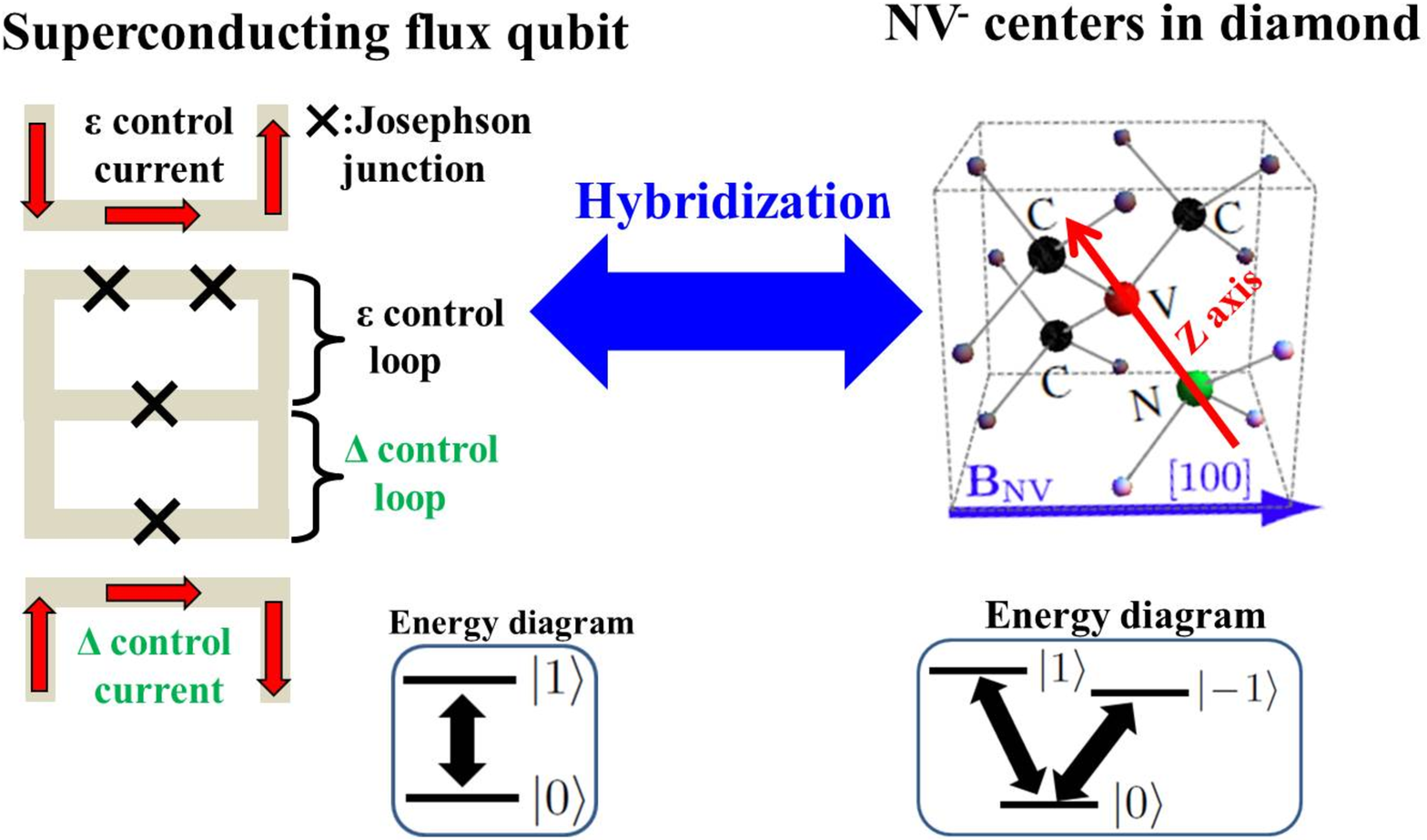}
\caption{Illustration of the hybrid system composed of a superconducting
 flux qubit and an ensemble of NV$^-$ centers. The flux qubit has four
 Josephson junctions that forms a two-level system. There are two control lines for the flux qubit. We use one of the control
 lines to change the energy bias $\epsilon $, and use the other line to
 change the energy gap $\Delta $. Diamond crystal is glued on top of a
 flux qubit, and this diamond contains NV$^-$ centers. The
 electrons spins trapped in the NV$^-$ center
 form a
 three-level system, and so we have a V-type energy level structure for the NV$^-$ center.
 }
\label{device}%
\end{figure}

Hybridization is a promising approach for quantum
computation \cite{wallquist2009hybridetal, xiang2013hybridetal}. Each system has
characteristic with its own advantages and disadvantages.
To couple different systems, we hope to pick up the advantage which each
system has.
One of the candidates for such hybrid systems is a superconducting
circuit such as a superconducting flux
qubit (FQ) and an electron spin ensemble such as nitrogen-vacancy (NV$^-$)
centers
\cite{imamouglu2009cavity,wesenberg2009quantumetal,schuster2010highetal,wu2010storageetal,kubo2010strongetal,amsuss2011cavityetal,kubo2011hybridetal,zhu2011coherent,kubo2012storageetal,kubo2012electronetal,marcos2010couplingetal,twamley2010superconducting,matsuzaki2012enhanced,saito2013towards,julsgaard2013quantumetal,diniz2011stronglyetal,zhudark2014,sandner2012strong}, 
as described in the Fig. \ref{device}.
High controllability of superconducting FQs
has already been achieved with existing technology \cite{ClarkeWilhelm01a}. Reliable gate
operations have been already demonstrated
\cite{bylander2011noiseetal}. Quantum non-demolition measurements can be performed by Josephson bifurcation
amplifier \cite{ClarkeWilhelm01a}. However,
despite significant effort, the coherence time of the FQ is
\textcolor{black}{of}
the order of $10$ $\mu$s \cite{bylander2011noiseetal,stern2014flux}.
On the other hand, the NV$^-$ center has a long coherence time \cite{balasubramanian2009ultralongetal,takahashi2008quenching,mizuochi2009coherence,kurucz2011spectroscopic,ishikawa2012optical,maurer2012room,bar2013solid}.
With dynamical decoupling, the coherence
time of electron of the NV$^-$ center is $0.6$ s
\cite{bar2013solid} that is much longer
than the FQ.
So, coupling the FQ with the NV$^-$ centers
is a promising way to obtain both controllability
and long coherence time \cite{imamouglu2009cavity,wesenberg2009quantumetal,schuster2010highetal,wu2010storageetal,kubo2010strongetal,amsuss2011cavityetal,kubo2011hybridetal,zhu2011coherent,kubo2012storageetal,kubo2012electronetal,marcos2010couplingetal,twamley2010superconducting,matsuzaki2012enhanced,saito2013towards,julsgaard2013quantumetal,diniz2011stronglyetal,zhudark2014,sandner2012strong}.

It is often useful
\textcolor{black}{to transfer the state} between the FQ and NV$^-$
centers for
quantum information processing.
We keep quantum states in the quantum memory (NV$^-$ centers) when gate operations
are not required.
On the
other hand, to perform gate operations, we need to transfer the quantum states
from the quantum memory to the quantum processor (FQ),
\textcolor{black}{which can be realized by using vacuum Rabi oscillation (VRO).}
However, \textcolor{black}{the error rate of
the state transfer in the current technology is an order of ten percent
\cite{kubo2011hybridetal,zhu2011coherent,saito2013towards}, which is too
large to perform quantum computation \cite{RHG01a,stephens2013high}.}
\textcolor{black}{The noise mainly comes from the
inhomogeneous broadening of the NV$^-$ centers.}
\textcolor{black}{Therefore, it is crucial to suppress the
decoherence of the NV$^-$ centers for computational tasks.}

\textcolor{black}{Improving the coherence time of the NV$^-$
center ensemble also has a fundamental importance in the area of quantum
metrology \cite{pham2011magnetic,hardal2013discrete},
quantum walk \cite{hardal2013discrete}, and quantum simulation
\cite{yang2012quantum}. In these applications, the efficiency strongly
depends on the coherence time of
the ensemble of NV$^-$ centers.
So it is essential in these areas to find a way to improve
the coherence time of the NV$^-$ centers.}

\textcolor{black}{
A cavity protection
\cite{steph2014nonmarkov,krimer2014non, diniz2011stronglyetal} is a
promising way to improve the coherence time of the NV$^-$ center ensemble. If the coupling
strength between the ensemble of NV$^-$ centers and a superconducting flux qubit (or
microwave cavity) is larger than the inhomogeneous width of the NV$^-$
centers, the collective mode of the NV$^-$ centers could be well decoupled
from the other sub-radiant states of the NV$^-$ so that inhomogeneous
broadening would be suppressed.
However, it has been shown that, if the spectral density of the
inhomogeneous broadening is described by a Lorentzian, the noise cannot be suppressed by the cavity protection
\cite{diniz2011stronglyetal,kurucz2011spectroscopic}.
Moreover, for the applications of quantum memory and quantum field
sensing, it is sometime necessary to turn off the interaction between the NV
centers and superconducting circuit. So it is better to have an
alternative scheme that will work even for such cases.
}

In this paper, we report an improvement of the lifetime of
the
\textcolor{black}{VRO}
by applying an in-plain magnetic field to this hybrid system.
We have observed \textcolor{black}{VRO}
with/without the magnetic field,
and the lifetime of the \textcolor{black}{VRO}
with magnetic field is
\textcolor{black}{nearly twice than}
without magnetic field. We have constructed a
theoretical model to reproduce these results, and have found that the
magnetic field suppress the inhomogeneous strain effect.

 \textcolor{black}{Let us now} describe our experimental set-up.
 Our system consists of a gap-tunable FQ \cite{paauw2009tuning,zhu2010coherent} on which
  a diamond
  crystal with NV$^-$ density of approximately $5\times 10^{17}$
  cm${}^{-3}$  is bonded \cite{saito2013towards}. \textcolor{black}{The NV$^-$ ensemble
  is created by the ion implantation
  and annealing in in vacuum \cite{zhu2011coherent,saito2013towards}.}
  The distance between the FQ and the surface of the diamond
 crystal is less than $1\mu m$. We can apply external magnetic field
 of 2.6 mT along the
 [100] crystalline axis \cite{saito2013towards}.
  \textcolor{black}{The gap-tunable FQ is fabricated with a
  superconducting loop containing four Josephson junctions. To control the FQ, a microwave line is fabricated around the FQ.
The FQ is designed to couple with the SQUID structure via magnetic fields.
}
\textcolor{black}{The probability of the excited state of the FQ is measured
 by
 the SQUID.}

\textcolor{black}{ To observe \textcolor{black}{the VRO,}
we perform
the following experiment.
First, we excited the FQ by applying microwave
  pulse where the FQ is decoupled from the
  NV$^-$ centers by the detuning. Second, we brought the FQ into the resonance of the
  NV$-$ centers by applying a magnetic flux.
   Finally, after a time $t$, we can measure the excited probability of the FQ
  via the SQUID. 
  The measurements
  were done
  in a dilution refrigerator at a temperature below 50 mK.}

We describe our system by the Hamiltonian
\cite{diniz2011stronglyetal,marcos2010couplingetal,kubo2011hybridetal,zhudark2014}
\begin{eqnarray}
 H=H_{\text{flux}}+H_{\text{int}}+H_{\text{ens}}\ \ \ \ \ \ \ \ \ \ \ \
  \ \ \ \ \ \ \ \ \ \ \ \ \  \\
 H_{\text{flux}}=\frac{\hbar}{2}\textcolor{black}{\Delta} \hat{\sigma }_z
  +\frac{\hbar}{2}\textcolor{black}{\epsilon}  \hat{\sigma}_x,\ \ \ \ \ \
  \ \ \ \ \ \ \ \ \ \ \ \ \ \ \ \ \ \ \\
  H_{\text{int}}=\hbar g_e\mu _B \hat{\sigma }_x \Big{(}\sum_{k=1}^{N}
   {\bf {B}}^{(k)}_{\text{qb}}\cdot {\bf{S}}_k\Big{)}
   \ \ \ \ \ \ \ \ \ \ \ \ \ \ \ \ \ \ \ \\
 H_{\text{ens}}=\Big{(}\sum_{k=1}^{N}\hbar D_{k}\hat{S}^2_{z,k}+\hbar
  E^{(k)}_1(\hat{S}^2_{x,k}-\hat{S}^2_{y,k})  \\
 +\hbar E^{(k)}_2(\hat{S}_{x,k}\hat{S}_{y,k}+\hat{S}_{y,k}\hat{S}
  _{x,k})+\hbar g_e\mu_B{\bf{B}}^{(k)}_{\text{NV}} \cdot {\bf{S_k}}
  \Big{)}
\end{eqnarray}
where $\hat{\sigma} _{x,y,z}$ denotes the Pauli matrix for FQ
with $\hat{\sigma }_x$ whose eigenstates corresponds to
two persistent-current states.
Also, we define $\hat{\sigma
}_{\pm}=\frac{1}{2}(\hat{\sigma }_x\pm i\hat{\sigma }_y)$.
The
electron spin 1 operators of the NV$^-$ center are represented by
$\hat{S}_{x,y,z}$. 
$H_{\text{flux}}$ denotes the Hamiltonian of the FQ where
$\Delta $ denotes the energy gap
and $\epsilon $ denotes the magnetic energy bias.  $ H_{\rm{ens}}$
represents the ensemble composed of $N$ individual NV$^-$ centers where
 $D/2\pi \simeq 2.878$GHz denotes a zero field splitting,  $E_k
=\sqrt{(E^{(k)}_1)^2+(E^{(k)}_2)^2}  $ denotes a strain induced
splitting, $g _e\mu_B{\bf{B}}_{\text{NV}} \cdot {\bf{S}}$ denotes a Zeeman
splitting, and
$\bf{B}_{\text{NV}}$ denotes  a magnetic field with $g_e \mu_B/2\pi
=28$MHz$\cdot $mT${}^{-1}$.
A quantization axis (z axis)
to be the direction from the vacancy to the nitrogen is set by the zero field
splitting of the NV$^-$ center. 
For a small magnetic field $D\gg g_e\mu_B |{\bf {B}}_{\text{NV}}|$,
  the x and y component of the magnetic field is
  insignificant to change
  quantized axis of the NV$^-$ center, and so we include only the effect
  of z axis of the
  field.
 We consider three relevant type of the magnetic field ${\bf{B}}_{\text{NV}}$: an in-plane external
 magnetic field $B_{\text{ex}}$, an inhomogeneous magnetic field
 due to P1 centers $B_{\text{inh}}$, and a hyperfine
 field from the nitrogen nuclear spins $B_{\text{hf}}$. 
The term $H_{\rm int}$ denotes the magnetic coupling
between the FQ and the NV$^-$ centers
where $ {\bf{B}}^{(k)}_{\text{qb}}$ represented the magnetic field
induced by persistent current of the
FQ.
Since collective enhancement of the coupling strength between the NV$^-$ and
FQ is not available along the z axis of the NV$^-$ center
\cite{marcos2010couplingetal}, we can ignore the coupling with $\hat{S}_z$.
\textcolor{black}{We} can write $H_{\text{int}}$ as $H_{\text{int}}=\hbar \sum_{k=1}^{N} g
\hat{\sigma }_x \cdot  (\hat{S}_{x,k}\cos \phi _k-\hat{S}_{y,k}\sin \phi
_k) $
where $g= g_e\mu _B B^{(xy)}_{\text{qb}}$ denotes a Zeeman splitting of
the NV$^-$ spin due to FQ magnetic field $B^{(xy)}_{\text{qb}}$ in
x-y plane and $\phi _k$ denotes the angle of the field in the plane.

For an ensemble of NV$^-$
\textcolor{black}{center with only a few excitations in it at most, we
can use the Holstein-Primakoff approximation to treat
NV${}^-$ spins as an ensemble of harmonic oscillators}
\cite{houdre1996vacuum}. We define
creation (destruction) operators \textcolor{black}{of a bright state and a dark state
of the NV center \cite{zhudark2014}} \textcolor{black}{by} $ \hat{b}_k^{\dagger}=|A_+\rangle
_k\langle 0|$ and $ \hat{d}_k^{\dagger}=|A_-\rangle _k  \langle 0|$, ($
\hat{b}_k=|0\rangle _k\langle A_+|$ and $ \hat{d}_k=|0\rangle _k
\langle A_-|$) where $|A_+\rangle _k= \frac{\cos \phi
_k}{\sqrt{2}}(|1\rangle_k+|-1\rangle _k)+ \frac{i\sin \phi
_k}{\sqrt{2}}(|1\rangle _k-|-1\rangle _k)$,
$|A_-\rangle _k=\frac{i\sin \phi _k}{\sqrt{2}}(|1\rangle_k+|-1\rangle
_k) +\frac{\cos \phi _k}{\sqrt{2}}(|1\rangle_k-|-1\rangle _k)$.
Here, $|A_+\rangle
_k\langle 0|$ denotes the state of the NV$^-$ center to be directly
coupled with the FQ while $|A_-\rangle
_k\langle 0|$ has no direct coupling with the FQ.
Since we
consider only one or zero excitation in a total system, we can also replace
a ladder operator of the FQ \textcolor{black}{with} a creation operator of a harmonic
oscillator as
$\hat{\sigma }_+ \rightarrow \hat{c}^{\dagger }$.

Moving to a rotating \textcolor{black}{frame with angular frequency
 $\omega $
 } defined by
 $U=e^{-i(\frac{1}{2}\omega \hat{\sigma }_z+\omega
 \hat{S}_z^2)t}$
 and making the rotating wave approximation, we obtain the simplified Hamiltonian
$H\simeq \hbar \omega _c\hat{c}^{\dagger }\hat{c} +\sum_{k=1}^{N}\Big{(} \hbar  \omega ^{(k)}_b\hat{b}_k^{\dagger
  }\hat{b}_k+\hbar \omega ^{(k)}_{d}\hat{d}_k^{\dagger}\hat{d}_k +\hbar
  g'
  (\hat{c}^{\dagger } \hat{b}_k+\hat{c}\hat{b}_k^{\dagger
  }) +\hbar (J_k+iJ'_k)\hat{b}_k^{\dagger }\hat{d}_k+\hbar (J_k-iJ'_k)\hat{b}_k\hat{d}_k^{\dagger }  \Big{)}$
where
\textcolor{black}{$\omega _c=\sqrt{\epsilon^2 +\Delta ^2}$}, $\omega ^{(k)}_b\simeq D_{k}-\omega -E_{y,k}$, $\omega
^{(k)}_d\simeq D_{k}-\omega +E_{y,k}$,
$J_k=g_e\mu _BB^{(k)}_z$,
 $J_k'=E_{x,k}$, 
$g'=g\Delta 
/\sqrt{\epsilon ^2+\Delta ^2}$,
$E_{x,k}=E_k\cos 2\phi _k$, and $E_{y,k}=E_k\sin
 2\phi_k$.

\begin{figure}[ht]
\includegraphics[scale=0.23]{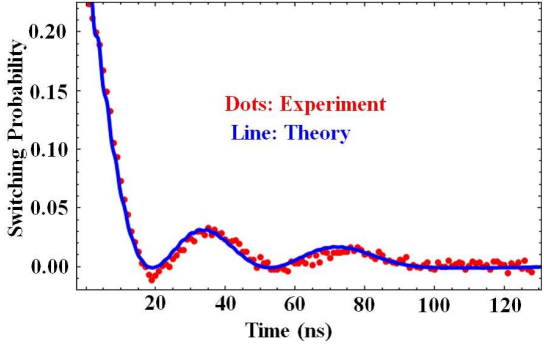}
\caption{Vacuum Rabi oscillations between the FQ and NV$^-$
 centers without an applied external magnetic field.
 Red dots show the experimental results while the blue line shows the results of a
numerical model where we use
 $N=1200$, $\epsilon =0$, $\Gamma _c/2\pi
 =0.3$MHz, $\Gamma_b/2\pi=\Gamma _d/2\pi=0.44$MHz, $\delta
 D_{k}/2\pi=0.08$MHz (FWHM), $\delta (g\mu_B B^{(k)}_z)/2\pi
 =3.1$MHz(FWHM), $\delta
 E_{1,k}/2\pi =\delta E_{2,k}/2\pi =4.4$MHz(FWHM), $A_{HF}=2.3$MHz, and $\sqrt{N}g/2\pi =13$MHz}
\label{vrabizerob}%
\end{figure}

\begin{figure}[ht]
\includegraphics[scale=0.23]{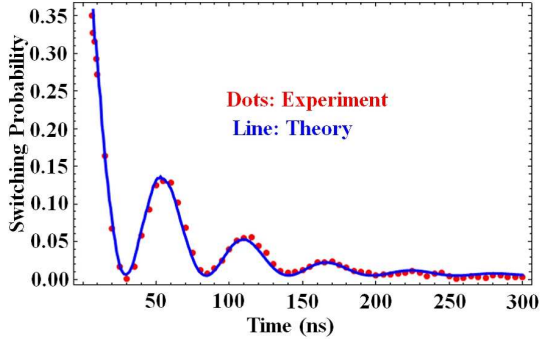}
\caption{Vacuum Rabi oscillations between the FQ and NV$^-$
 centers with an applied external magnetic field of $2.6$ mT along the
[100] crystalline axis. Except the
 magnetic field, we use the
 same parameters as those in
 Fig. \ref{vrabizerob}. }
\label{vrabib}%
\end{figure}

 \textcolor{black}{Now the dynamics of this hybrid system can be investigated using the Heisenberg equations of motions.}
We
write
Heisenberg equations of motion as
 \begin{eqnarray}
 \frac{d}{dt}\hat{c}=-i\omega _c \hat{c}-i (\sum_{k=1}^{N}g_k\sin \xi
  \cdot \hat{b}_k)-\Gamma_c\hat{c}\ \ \ \ \ \ \ \ \ 
  \\
 \frac{d}{dt}\hat{b}_k=-i\omega ^{(k)}_b \hat{b}_k-iJ_k\hat{d}_k+J'
  _k\hat{d}_k-ig \sin \xi _k\cdot \hat{c} -\Gamma _b\hat{b}_k
  \\
 \frac{d}{dt}\hat{d}_k=-i\omega ^{(k)}_d \hat{d}_k-iJ_k\hat{b}_k- J'
  _k\hat{b}_k-\Gamma _d\hat{d}_k
  \ \ \ \ \ \ \ \ \ 
 \end{eqnarray}
where $\Gamma _c$, $\Gamma _b$, and $\Gamma _d$ denote the decay rate
of
$\hat{c}$, $\hat{b}$, and $\hat{d}$,
respectively.
We numerically solve these equations with an initial state
of $|\psi (t=0)\rangle =\hat{c}^{\dagger
}|\text{vac}\rangle $,
and plot
 the renormalized excitation probability of
the FQ (which corresponds a switching probability of SQUID)
  in the Fig. \ref{vrabizerob} and
Fig. \ref{vrabib}.


\textcolor{black}{Now let us detail our simulation technique and the core elements of it.}
The Lorentzian distributions, \textcolor{black}{which have been typically used to
describe the inhomogeneous broadening of the NV$^-$ centers \cite{zhu2011coherent,zhudark2014,kubo2011hybridetal,kubo2012electronetal}}, are assumed for $D^{(k)}_0$, $E_{1,k}$, and $E_{2,k}$
$(k=1,2,\cdots ,N)$ to include the effect of the inhomogeneous
 lattice distortion of the NV$^-$ centers.
 Next, due to the electron spin-half bath in the environment such as P1 center,
 randomized magnetic field on the NV center exists, and the nitrogen nuclear spin
splits the electron-spin energy into three level via a hyperfine
coupling. To include both these two effect,
we use a random distribution of the magnetic fields with
the form of
the mixture of three Lorentzian functions that are separated with $2\pi \times
2.3 $ MHz due to the hyperfine interaction with ${}^{14}N$ nuclear spin \cite{kubo2011hybridetal, saito2013towards}.
With these assumptions, we have reproduced the \textcolor{black}{VRO}
with/without applied magnetic field in the Fig. \ref{vrabizerob} and
Fig. \ref{vrabib}. \textcolor{black}{While the FQ can induce the
 transition both $|0\rangle _k \leftrightarrow |-1\rangle _k$ and
 $|0\rangle _k \leftrightarrow |1\rangle _k$ with zero applied magnetic
 field, the FQ induces only one of them with applied magnetic field
 of a few mT due to the detuning effect
 \cite{kubo2010strongetal,zhu2011coherent,marcos2010couplingetal,saito2013towards,diniz2011stronglyetal}. This
 changes the effective coupling strength
 between the FQ and NV centers, and is the cause of the
 different time interval of the oscillations in the Fig. \ref{vrabizerob} and
Fig. \ref{vrabib}.}
With \textcolor{black}{an applied magnetic field of 2.6
mT},
the \textcolor{black}{VRO}
can be observed until around $170$ ns while we cannot observe a clear
oscillation \textcolor{black}{beyond} $100$ ns without \textcolor{black}{an}
applied magnetic field.
 \textcolor{black}{Thus, it} \textcolor{black}{shows} the external
magnetic field improves the lifetime of the
\textcolor{black}{VRO}.

\textcolor{black}{We explain how we determine the parameters for the
numerical simulations.
$\Gamma _c$
can be determined by the $T_1$ measurement of the FQ, which was
performed independent of the VRO experiment \cite{saito2013towards}. Since the
frequency shift of $E$ is 50 times larger than that of $D$ when an
electric field is applied \cite{dolde2011electricetal}, we use $\delta E\simeq 50
\delta D$ in this paper. 
It is known that,
from the spectroscopic measurements, a sharp peak 
located in the middle of the avoided crossing 
structure was observed in this hybrid system \cite{zhudark2014}, and one can determine the $\Gamma _b(=\Gamma _d)$ from
the width of this sharp peak \cite{zhudark2014}. The time period of the
VRO let us specify the value of $\sqrt{N}g$. Moreover, as we
describe later, the envelope of the VRO with (without) magnetic field is
mainly determined by $\delta B_z$ ($\delta B_z$ and
$\delta E$). So, by fitting the spectroscopy and VRO with/without magnetic field, one can specify necessary
parameters for our model.}

\textcolor{black}{We understand why}
the applied magnetic field actually
improves the coherence time of the NV$^-$ centers.
 There are two relevant \textcolor{black}{decoherence source for the NV centers,
 inhomogeneous magnetic
 fields and the strain distribution \cite{zhu2011coherent,kubo2011hybridetal, saito2013towards}.}
The magnetic-field noise
comes from an effective Hamiltonian between P1 center and NV center as
$H_{\text{eff}}=\sum_{k,l}\lambda _{k,l}\hat{\sigma }^{(\text{P1},l)}_z
\hat{S}^{(k)}_z$
where flip-flop term is negligible due to the large energy difference
between them. However, this term commutes with the Zeeman term
of the applied external magnetic field as $g _e\mu
_BB_{\text{ex}}\hat{S}_z$, and so external magnetic field cannot
affect this.
On the other hand,
\textcolor{black}{the Hamiltonian of the 
inhomogeneous strain} does not commute with the Zeeman Hamiltonian
of the applied external magnetic field.
\textcolor{black}{Actually, the Hamiltonian of the $k$ th NV$^-$ center is written
as $H^{(k)}_{\text{NV}}=\hbar \omega ^{(k)}_b \hat{b}_k^{\dagger }\hat{b}_k+\hbar
\omega ^{(k)}_d \hat{d}_k^{\dagger }\hat{d}_k+
\big{(}(J_k+iJ'_k)\hat{b}^{\dagger }_k\hat{d}_k+\text{e.c}\big{)}$, and
the eigenenergies are $E_0=0$ and $E_{\pm }=D\pm
\sqrt{E^2_{x,k}+E^2_{y,k}+(g_e\mu _BB_z^{(k)})^2}$. If the magnetic
field is large, we can expand the eigenenergies of the excited states as
$E_+\simeq g_e \mu _B B^{(k)}_z\pm \frac{E^2_{x,k}+E^2_{y,k}}{2g_e \mu
_B B^{(k)}_z}$. So the effect of the variations of $E_{x,k}$ and
$E_{y,k}$ becomes negligible, and this could improve the lifetime of the
VRO}
\textcolor{black}{\footnote{Strictly speaking, there is a hyperfine
coupling of $2.3$ MHz even without applying magnetic field, which can be
considered as an effective magnetic field from the Nitrogen nuclear
spins. However, in our sample, the strain distribution is larger than
the hyperfine coupling, and so the hyperfine coupling cannot
significantly suppress
the strain variations.}}.

\begin{figure}[ht]
\includegraphics[scale=0.19]{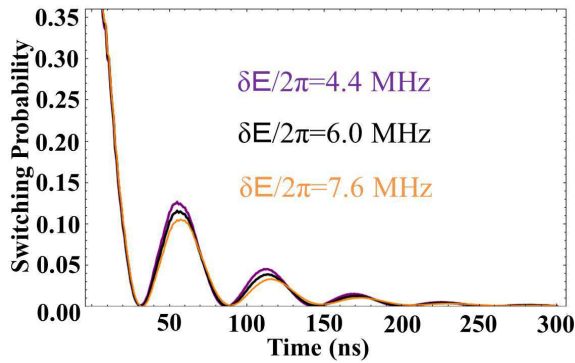}
\caption{Numerical simulations of vacuum Rabi oscillations with the applied magnetic field for
 several values of the strains. Except the
 values of the strains, we use the
 same parameters as those in
 Fig. \ref{vrabib}. }
\label{strainchange}%
\end{figure}

\textcolor{black}{To confirm this effect, we performed another numerical
simulation of the VRO with applied magnetic field for the strain values of $\delta E/2\pi =4.4, 6.0, 7.6$ MHz in the
Fig. \ref{strainchange}. Interestingly, the three VROs shown
in the Fig. \ref{strainchange} are almost the same. These results clearly show that the lifetime of
the VRO is quite robust against the inhomogeneous strains.}
So
we conclude that the external magnetic field suppress
the effect of inhomogeneous strain so that the improvement of the lifetime has been observed in our experiment.
\textcolor{black}{It is worth mentioning that, although possible
suppression of
the strain distribution by the applied magnetic field is mentioned in
\cite{acosta2013optical}, we firstly 
demonstrate such a mechanism by the experiment.}

\textcolor{black}{It is known that applying
transversal magnetic field can improve the coherence time of the NV$^-$
centers
when the strain distribution is much larger than the decoherence rate
due to the randomized magnetic fields
\cite{shin2013suppression}. Here, the application of the transversal
magnetic field can suppress the decoherence due to the environmental magnetic field
while this cannot suppress the inhomogeneous broadening of the strain \cite{shin2013suppression}.
On the other hand, our scheme to apply horizontal magnetic field is complemental to
this, because we can suppress the strain inhomogeneous broadening. Our scheme has an
advantage
 especially when we can decrease the broadening due to the
magnetic field by using another technique.
}

\textcolor{black}{Actually, it is possible to combine our scheme with another
technique to reduce the inhomogeneous magnetic fields. For example,
one way to suppress the magnetic noise for the NV$^-$ centers is to
decrease the P1 centers (a nitrogen
atom substituting a carbon atom) by
using differently synthesized diamond crystals \cite{saito2013towards}. P1 centers are
considered as a
electron spin-half bath, and they cause randomized magnetic field to decoher
the NV$^-$ centers \cite{kubo2011hybridetal, zhu2011coherent}. However,
 reduction of the P1 centers cannot suppress the noise due to the
 strain. Therefore, by applying a magnetic field with a diamond
 having less P1 centers, further improvement of the coherence time
 should be possible. This provides us with a sensitive diamond-based
 field sensor, and will be useful for a long-lived quantum
 memory during  quantum computation as future applications.}

In conclusion, we have observed an improvement of a lifetime of the
vacuum Rabi \textcolor{black}{oscillations} between the FQ and NV centers by
applying an in-plain magnetic field. By reproducing the experimental
result from a theoretical model, we have found that the applied magnetic
field can suppress the inhomogeneous broadening of the strain.
This result is a relevant step toward to the realization of the
long-lived quantum memory for a superconducting flux qubit.

 We thank R. Ams\"uss and H. Nakano for valuable discussions.
 This work \textcolor{black}{was supported by KAKENHI(S) 25220601}
 and in part
by
Commissioned Research of NICT.


\end{document}